\documentclass[a4paper,fleqn]{cas-sc}
\usepackage{soul}
\usepackage[sort&compress,numbers]{natbib}
\usepackage{lineno}
\usepackage{setspace}
\usepackage{graphics}
\usepackage{graphicx}
\usepackage{cuted}
\usepackage{lipsum}

\def\tsc#1{\csdef{#1}{\textsc{\lowercase{#1}}\xspace}}
\tsc{WGM}
\tsc{QE}
\tsc{EP}
\tsc{PMS}
\tsc{BEC}
\tsc{DE}

\begin{document}

\let\WriteBookmarks\relax
\def\floatpagepagefraction{1}
\def\textpagefraction{.001}
\shorttitle{O$_2$ Adsorption on Defective Penta-Graphene Lattices}
\shortauthors{Lima \textit{et~al}.}

\title [mode = title]{O$_2$ Adsorption on Defective Penta-Graphene Lattices: A DFT Study}

\author[1]{Kleuton A. Lopes Lima}
\author[1]{Marcelo L. Pereira J\'unior}
\author[1]{F\'abio F. Monteiro}
\author[1]{Luiz F. Roncaratti J\'unior}
\author[1]{Luiz A. Ribeiro J\'unior}
\cormark[1]
\ead{ribeirojr@unb.br}

\address[1]{Institute of Physics, University of Bras\'ilia, 70910-900, Bras\'ilia, Brazil}

\cortext[cor1]{Corresponding author}

\begin{abstract}
Penta-Graphene (PG) was theoretically proposed as a new carbon allotrope with a 2D structure. PG has revealed interesting gas sensing properties. Here, the structural and electronic properties of defective PG lattices interacting with an oxygen molecule were theoretically studied by employing density functional theory calculations. Results show that PG lattices with a $sp^3$-like single-atom vacancy presented higher adsorption energy than the $sp^2$-like one. Remarkably, PG lattices with a $sp^3$-like defect presented a clear degree of selectivity for the molecule orientation by changing their bandgap configurations. Importantly, the adsorption energies were obtained using the improved Lennard-Jones (ILJ) potential.  
\end{abstract}


\maketitle
\doublespacing

\section{Introduction}
\label{sec1}

In the last two decades, the need for environmental, industrial, and biological monitoring of O$_2$ concentration has stimulated the growing interest in developing new oxygen sensor technologies \cite{tyson_CC,wang_nanoscale,stoeckel_AM}. Carbon-based 3D and 2D nanomaterials are known as suitable sensors of small molecules such as O$_2$, CO, NO, and NO$_2$ with the potential of monitoring low concentrations of these gases, presenting optimal response times \cite{leenaerts_PRB,schedin_NM,hu_PCCP,yuan_AM,bai_carbon,khavrus_PE,valencia_JPCC}. Particularly, 2D structures of these nanomaterials have been both experimentally and theoretically studied regarding their potential of acting as gas sensors, mostly due to their large surface area and high carrier mobility \cite{chen_SABC,zhu_PRL,meng_IEEE}. In this sense, it was reported recently that the electronic properties of 2D nanomaterials, such as graphene and MoS$_2$, are altered upon adsorption of small molecules \cite{cho_SR,ou_ACSnano}.

Pristine graphene presents a stable $sp^2$-like hybridization of carbon bonds and null bandgap \cite{blake_APL,nair_Science,neto_RMP}. These features make it inefficient for gas adsorption and, therefore, not suitable for developing gas sensor devices. On the other hand, Penta-Graphene (PG) --- a new 2D carbon allotrope with a Cairo tessellation based lattice (pentagonal arrangement of atoms) \cite{zhang_pnas,ewels_2015} --- was theoretically proposed as a structure that contains both $sp^3$-like and $sp^2$-like hybridizations of carbon bonds, which are more interesting when it comes to open new channels for the gas adsorption mechanism \cite{stuve_JPC,coffman_APL}. Due to the tetrahedral character of the $sp^3$-like hybridization, the PG surface is not precisely flat, which suggests the existence of regions with a higher chance for the adsorption of molecules. This particular feature also broadens the options to use PG  as an active layer in novel sensor prototypes \cite{zhang_pnas}. Although PG has shown promising trends to develop gas sensor applications \cite{qin_nanoscale}, due to its unavailability of synthesis, investigations in this direction are still scarce. 

Some theoretical contributions in the literature, mostly based on the density functional theory (DFT) calculations, have investigated the adsorption mechanism of small molecules in PG \cite{chen_SMT,chen_SMT,cheng_nanoscale,li_RSCadvances,qin_nanoscale,manjanath_2dmaterials,Yen_PO,feng_2017,enriquez_IJHE}. Generally, the results have revealed the existence of substantial adsorption energies for the complex molecule/PG with moderate charge transfer for small molecules such as H$_2$O, H$_2$S, NH$_3$, SO$_2$, and NO \cite{cheng_nanoscale}. When it comes to the adsorption mechanism of oxygen molecules on PG lattices, studies in the literature are very few \cite{li_RSCadvances}. It is well known that the presence of lattice defects is inevitable during the manufacturing process of nanomaterials \cite{li_NC,jeevanandam_BJN,mudunkotuwa_JEM}. In this sense, investigations that take into account the impact of single-atom vacancies on the interaction between small molecules and PG membranes can contribute to gain a broader understanding of the adsorption mechanism in systems with carbon-based active layers.

Herein, we employed DFT calculations to numerically study the effect of O$_2$ adsorption on the electronic and structural properties of PG lattices endowed of single-atom vacancies. Remarkably, our results point to the possibility of O$_2$ adsorption on PG at room temperature with reasonable adsorption energy, low recovering time, and a good degree of selectivity. The calculations performed here suggest that PG can be a promising candidate for the production of O$_2$ sensors and open a channel for the understanding of the adsorption mechanism of small molecules in carbon-based lattices.

\section{Methodology}
\label{sec2}

\begin{figure*}[pos=ht]
	\centering
	\includegraphics[width=0.8\linewidth]{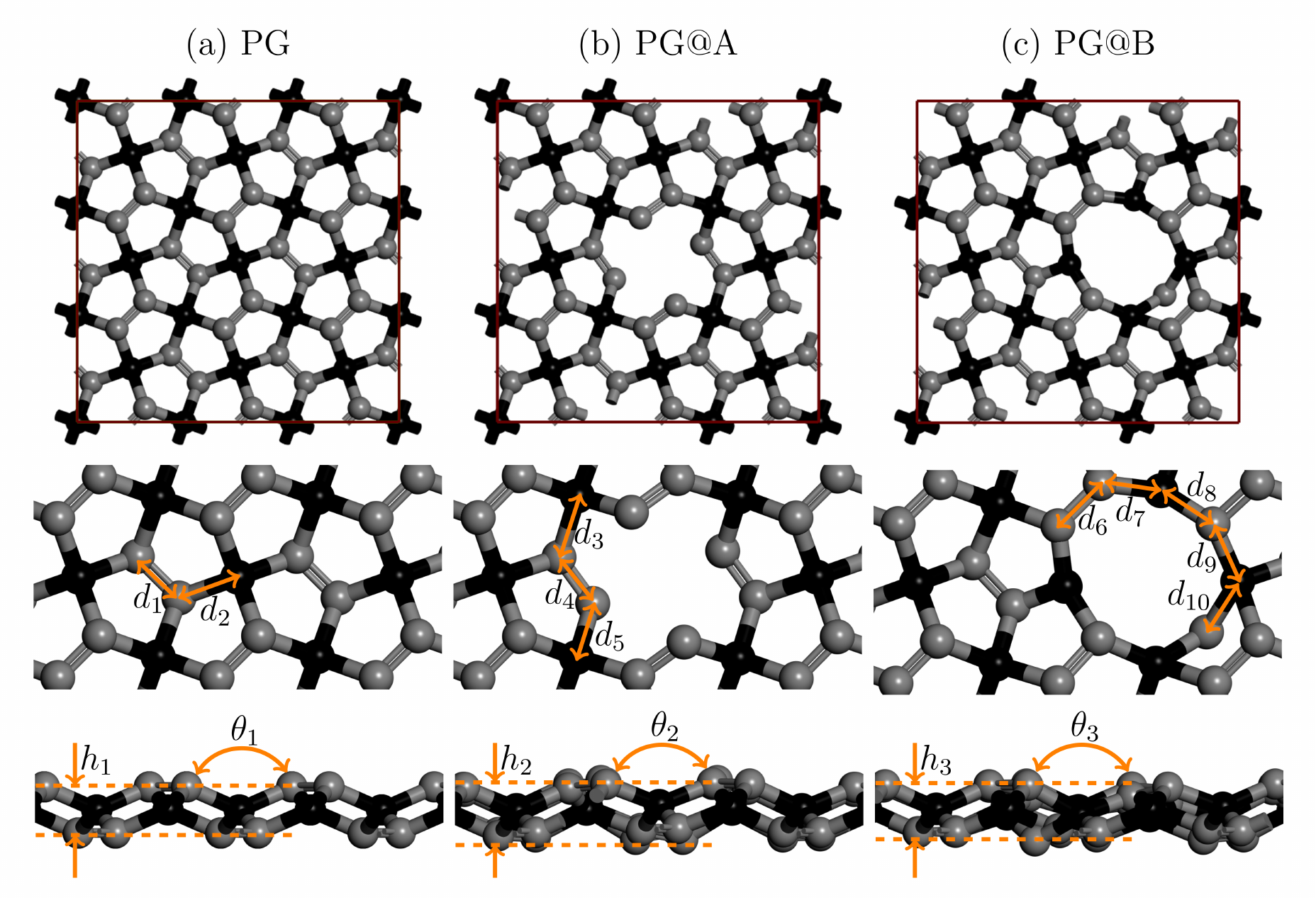}
	\caption{Top panels: Schematic representation for the ground state configurations of (a) non-defective PG, (b) a PG lattice with monovacancy defect at a $sp^3$-hybridized carbon atom, and (c) a PG lattice with monovacancy defect at a $sp^2$-hybridized carbon atom. Middle panels: enlarged regions to highlight the lattice defects. Bottom panels: side view of the PG lattices. In the color scheme, the $sp^3$-hybridized carbon atoms are the black spheres and $sp^2$-hybridized carbon atoms are the gray ones.}
	\label{fig:structures}
\end{figure*}

The structural and electronic properties of PG/O$_2$ complexes were studied using the DMol$^3$ module as implemented in Biovia Materials Studio software \cite{delley_JCP,delley_JCP2,andzelm_JCP}. In all calculations, the Local Density Approximation (LDA) is considered employing the Perdew-Wang (PWC) functional with unrestricted spin (DNP), and a numerical basis set of atomic orbitals with polarized functions \cite{perdew_PRB,kohn_PR}. The BSSE correction is used through the counterpoise method, and the nuclei-valence electron interactions are represented by the inclusion of semi-core DFT pseudopotentials \cite{delley_PRB}. The K points in the Brillouin zone are considered within a $14\times14\times1$ Monkhorst-Pack mesh \cite{monkhorst_PRB}. Ground state structures for the PG lattices, as presented in Figure \ref{fig:structures}, are obtained by defining the following tolerances: $1\times10^{-5}$ for the self-consistent field, 0.002 Ha/\r{A} for maximum force, and 0.005 \r{A} for maximum displacement. A $3\times3$ supercell with a vacuum space of 30 \r{A} is used to model the interacting PG/O2 complexes. It is worthwhile to stress that this set of parameters was successfully used in other theoretical studies, where the adsorption of small molecules on the surface of nanostructures was also investigated \cite{lima_jmm,paura_jpca,paura_rsc,lima_njc}.

In Figure \ref{fig:structures}, the PG lattices contain $sp^3$-hybridized carbon atoms (black spheres) and $sp^2$-hybridized carbon atoms (gray spheres). Figure \ref{fig:structures}(a) presents the ground state structure for a non-defective lattice (PG). Figures \ref{fig:structures}(b) and \ref{fig:structures}(c) depict the ground state structures for PG lattices with a monovacancy defect at $sp^3$-hybridized (PG@A) and $sp^2$-hybridized carbon (PG@B) atoms, respectively. It is worthwhile to stress that these structures were studied very recently \cite{manjanath_2dmaterials}. The structural parameters obtained here (highlighted in Figure \ref{fig:structures}) are $d$ (in-plane C--C distance ), $h$ (out-of-plane C--C distance), and $\theta$ (planarity degree). Their values are: $d_1=1.34$ \r{A}, $d_2=1.54$ \r{A}, $d_3=1.55$ r{A}, $d_4=1.33$ \r{A}, $d_5=1.50$ \r{A}, $d_6=1.37$ \r{A}, $d_7=1.41$ \r{A}, $d_8=1.45$ \r{A}, $d_9=1.51$ \r{A}, $d_{10}=1.51$ \r{A}, $h_1=0.039$ \r{A}, $h_2=0.050$ \r{A}, $h_3=0.047$ \r{A}, $\theta_1=135.01^\circ$, $\theta_2=139.21^\circ$, and $\theta_3=137.63^\circ$. Importantly, these values are in good agreement with the ones reported in other theoretical studies in literature \cite{qin_nanoscale,manjanath_2dmaterials}. 
 
\begin{figure}[pos=ht]
\centering
\includegraphics[width=0.35\linewidth]{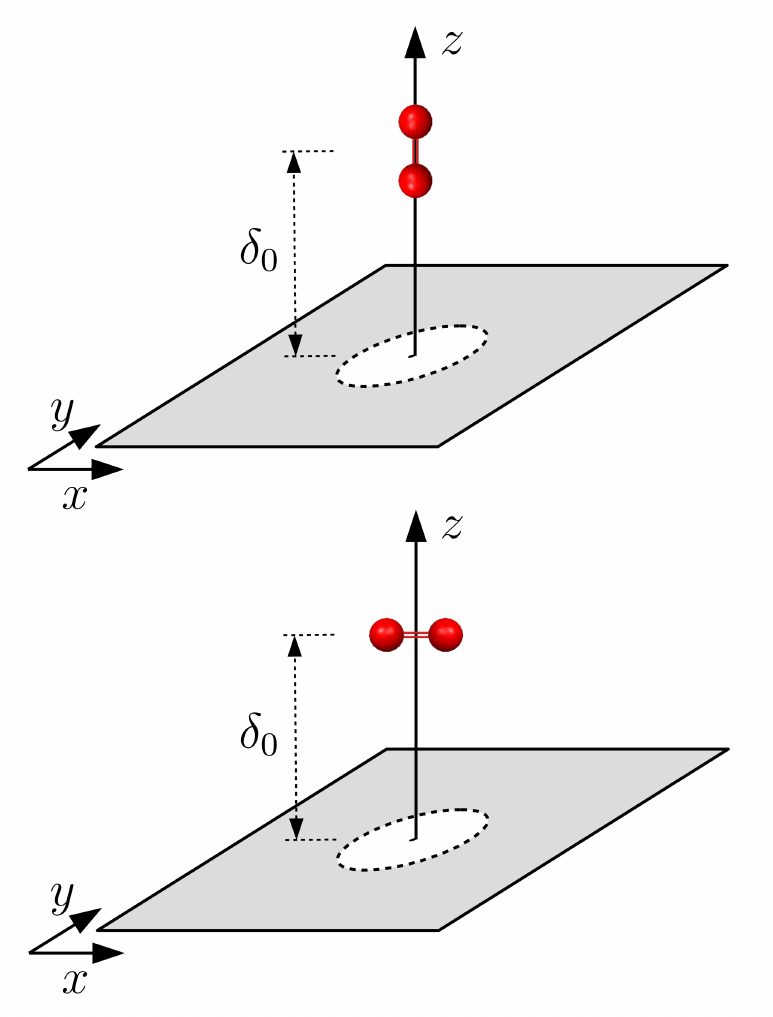}
\caption{Schematic representation of the computational approach (initial system configurations) used here to obtain the adsorption energy curves for the O$_2$ interaction with all the PG lattices presented in Figure \ref{fig:structures}. The $O_2$ molecule is initially positioned 7 \r{A} above the PG plane.}
\label{fig:protocol}
\end{figure}

In our computational approach, the O$_2$ molecule is positioned parallel (O$_2$-H) or perpendicular (O$_2$-V) to the PG plane, at an initial distance of 7.0 \r{A} (see Figure \ref{fig:protocol}). The adsorption mechanism for these PG/O2 complexes was investigated considering six cases, where the O$_2$ molecule approaches the PG, PG@A, and PG@B surfaces, forming the following complexes: PG/O$_2$-H, PG/O$_2$-V, PG@A/O$_2$-H, PG@A/O$_2$-V, PG@B/O$_2$-H, and PG@B/O$_2$-V. In Figure \ref{fig:protocol}, $\delta_0$ represents the distance between the centroids of the molecule and the PG lattice. By varying $\delta_0$, the O$_2$ moved towards the PG plane and we obtained adsorption energy ($E_{ads}$) curves using the expression: $E_{ads} = E_{(PG+O_2)} - E_{PG} - E_{O_2}$, where  $E_{PG}$, $E_{O_2}$, and $E_{(PG+O2)}$ are the total energies for isolated PG, isolated O$_2$, and O$_2$ adsorbed on PG surface, respectively. For a more detailed description of the PG/O$_2$ adsorption mechanism, the case of higher adsorption energy is further investigated through its electronic properties. This case was identified through the adsorption energy curves. These curves were fitted using the Improved Lennard-Jones (ILJ) equation \cite{pirani_PCCP}.

\section{Results}
\label{sec3}

\begin{figure}[pos=ht]
	\centering
	\includegraphics[width=0.65\linewidth]{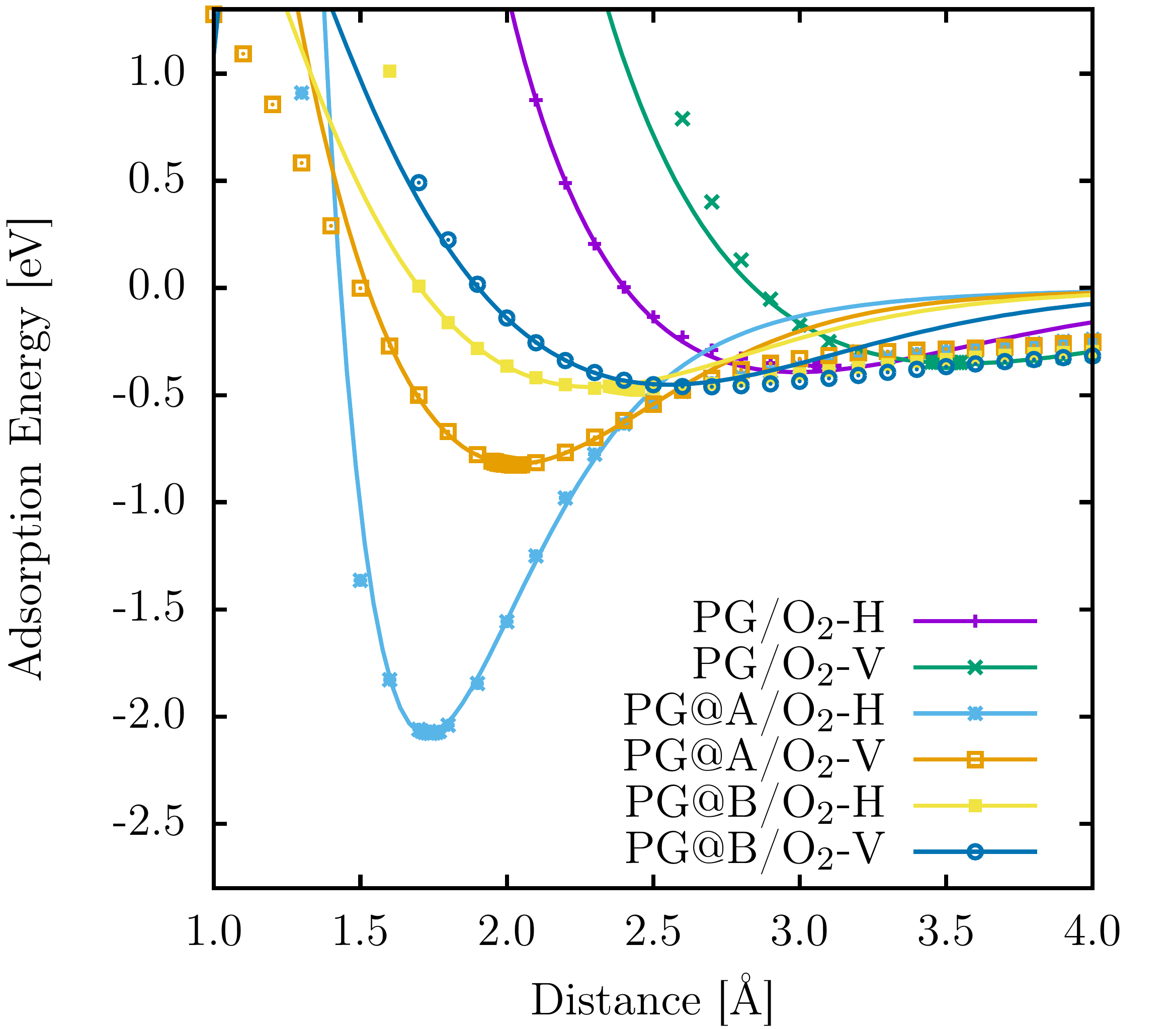}
	\caption{Adsorption energy curves for all the PG/$O_2$ complexes studied here. The curves were fitted using the ILJ equation \cite{pirani_PCCP}. The initial configurations are illustrated in Figure \ref{fig:protocol}.}
	\label{fig:curves}
\end{figure}

We begin our discussions by presenting the adsorption energy curves that were obtained using the protocol discussed above. Figure \ref{fig:curves} displays these curves and their related ILJ fitting \cite{pirani_PCCP}. In this figure, one can note that the interplay between the adsorption energy and the distance between O$_2$ and PG yields typical potential energy curves. The PG/$O_2$-H, PG/O$_2$-V, PG@B/O$_2$-H, and PG@B/O$_2$-V showed a physisorption mechanism since their curves dot not present a clearly defined potential well. In these systems, the electronic properties of the adsorbent are slightly changed in the presence of an adsorbate (as discussed later). The obtained adsorption energies ($\epsilon$ parameter in reference \cite{pirani_PCCP}) and equilibrium distances ($r_m$) are presented in Table \ref{tab:adsorption}. These values for PG/$O_2$-H, PG/O$_2$-V, PG@B/O$_2$-H, and PG@B/O$_2$-V cases are similar to the ones obtained in other theoretical studies in which the adsorption of an oxygen molecule on pristine boron-nitrite and graphene membranes were considered \cite{yan_JAP,sun_nanoscale}. A different adsorption mechanism is obtained for PG lattices endowed with a $sp^3$-type vacancy. In Figure \ref{fig:curves}, one can note that the PG@A/O$_2$-H e PG@A/O$_2$-V cases present higher reactivity (or higher adsorption energies) than the other cases. The adsorption energy for the PG@A/O$_2$-H case is, at least, twice higher than all the other cases. Particularly, the adsorption process in PG@A/O$_2$-H system denotes a chemisorption mechanism, and the electronic properties of the PG are affected by its interaction with the O$_2$ molecule, as it will be discussed later. In contrast with the PG and PG@B cases, the PG@A/O$_2$-H e PG@A/O$_2$-V curves present a potential well. The vertical orientation of the $O_2$ molecule, regarding the PG plane, leads the two oxygen atoms to interact almost in the same fashion with the PG@A sheet. This trend is not observed in the PG@A/O$_2$-V case in which only one of the oxygen atoms is positioned close to the PG@A surface. Moreover, in the PG@A lattice, the atoms within the defective region are closer to each other than in the PG@B one. This feature is due to a tendency of the carbon atoms in forming new bonds in the vicinity of the vacancy, which reduces the area of the defective region and increases the electrostatic potential. 

\begin{table}[pos=ht]
\begin{tabular}{|c|c|c|c|}
\hline
\textbf{System} & \textbf{$ \epsilon $ [eV]} & \textbf{$ r_m $ [\r{A}]} \\ \hline
PG/O$_2$-H      & 0.39                  & 2.99             \\ \hline
PG/O$_2$-V      & 0.35                  & 3.55             \\ \hline
PG@A/O$_2$-H    & 2.08                  & 1.72             \\ \hline
PG@A/O$_2$-V    & 0.82                  & 2.03             \\ \hline
PG@B/O$_2$-H    & 0.46                  & 2.29             \\ \hline
PG@B/O$_2$-V    & -0.53                 & 2.55             \\ \hline
\end{tabular}
\caption{Calculated adsorption energies $\epsilon$ and equilibrium distances $r_m$ from the curve fitting presented in Figure \ref{fig:curves}, which was performed through the ILJ equation \cite{pirani_PCCP}.}
\label{tab:adsorption}
\end{table}

\begin{figure}[pos=ht]
	\centering
	\includegraphics[width=0.7\linewidth]{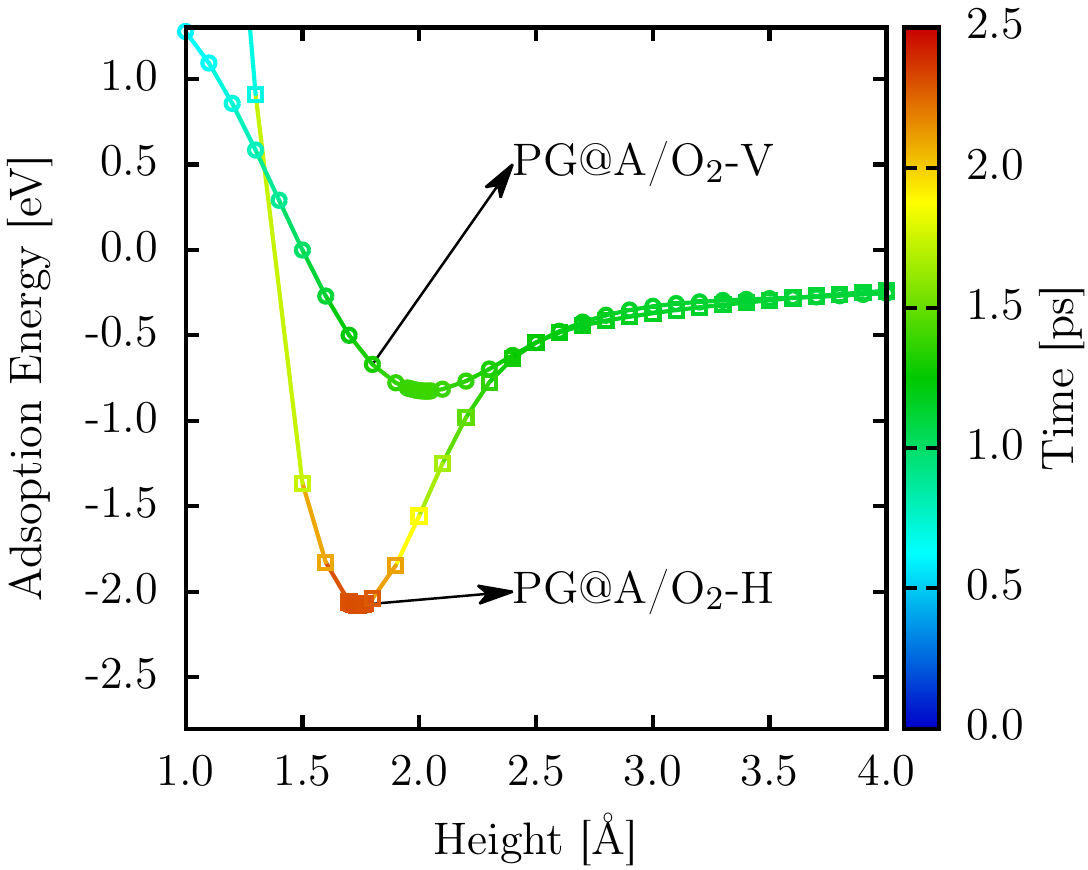}
	\caption{Recovering time values for PG@A/O$_2$-H and PG@A/O$_2$-V adsorption energy curves, depicted in Figure \ref{fig:curves}. The recovering time values are presented in the color palette.}
	\label{fig:recover}
\end{figure}

Now, we analyze the recovering time ($\tau$) \cite{timsorn_IOP} --- that corresponds to the transient time in which the molecule adsorption takes place --- for the systems with better adsorption performances (PG@A/O$_2$-H e PG@A/O$_2$-V). The recovering time is calculated as $\tau=1\nu \times exp(-E_{ads}/k_BT)$, where $\nu$ is the molecule oscillation frequency ($10^{12}s^{-1}$ \cite{peng_CPL}), $k_B$ is the  Boltzmann constant, and $T$ the temperature (298 K). In Figure \ref{fig:recover}, the recovering time values for PG@A/O$_2$-H and PG@A/O$_2$-V adsorption energy curves are presented in the color palette. In this way, one can note that the recovering time varies from 0.5 ps up to 2.5 ps. For the lowest absorption energy (PG@A/O$_2$-H case) $\tau=2.31$ ps whereas for the PG@A/O$_2$-V one, the highest recovering time is about 1.5 ps. It is worthwhile to stress that the transient times obtained for the O$_2$ adsorption in the PG@A lattice are large enough to realize the interaction between them, which changes the electronic properties of the PG (as discussed below), before the molecule diffusion.

\begin{figure*}[pos=ht]
	\centering
	\includegraphics[width=\linewidth]{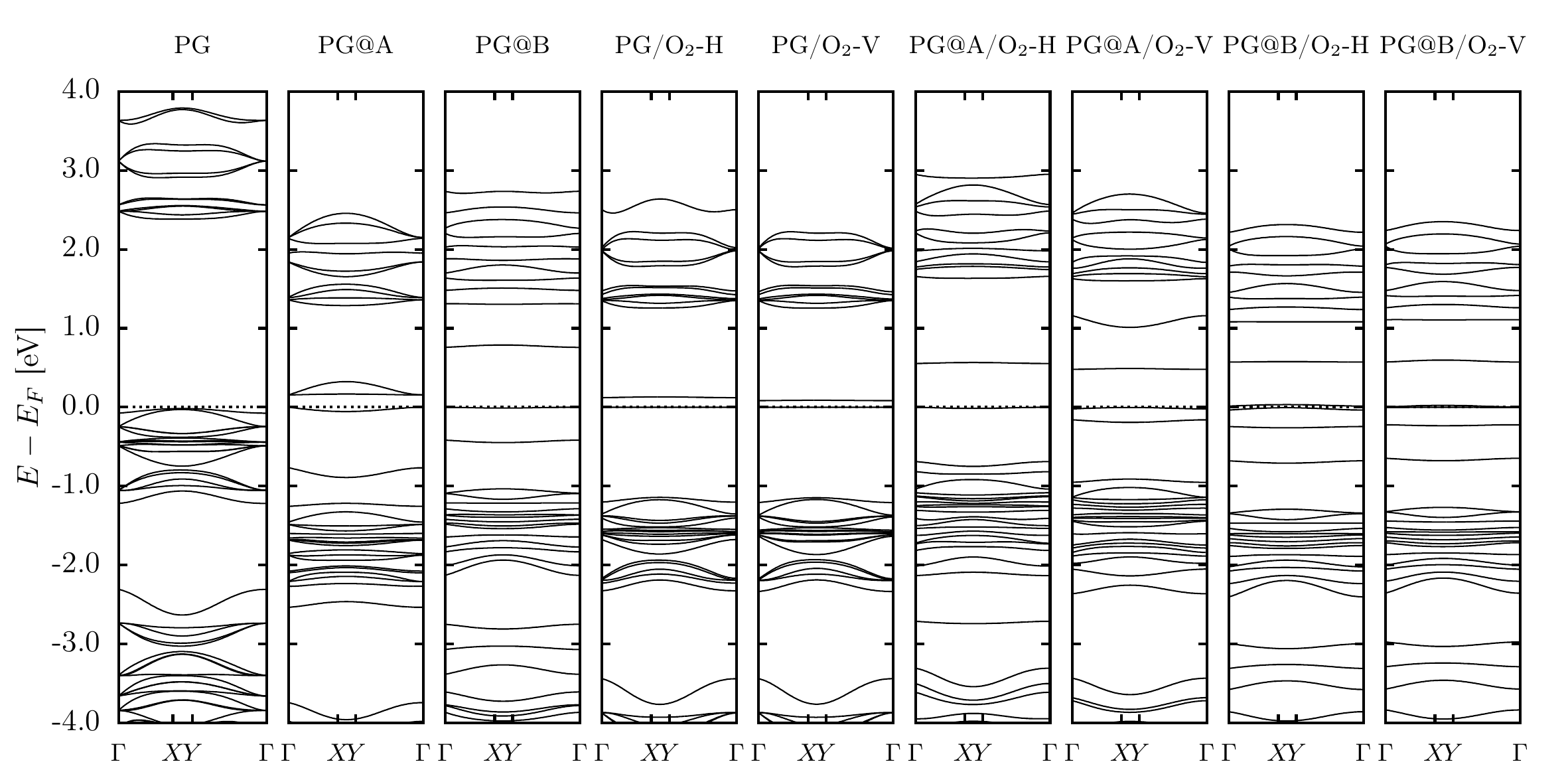}
	\caption{Electronic band structures for the systems investigated.}
	\label{fig:bandgap}
\end{figure*}

The electronic band structures for the systems investigated here are presented in Figure \ref{fig:bandgap}. The left-most panel illustrates that the non-defective PG presents a quasi-direct bandgap of about 2.35 eV, which is in good agreement with the values reported in the references \cite{enriquez_IJHE,zhang_pnas,qin_nanoscale}. As expected, the inclusion of a single-atom vacancy leads to the appearance of energy levels within the bandgap. Similar band structure signatures for the PG@A and PG@B cases were reported in other DFT-based studies \cite{enriquez_IJHE}. Upon adsorption, the valence and conduction bands of non-defective PG suffer an energy shifting (about -1.1 eV) for the PG/O$_2$-H and PG/O$_2$-V cases, preserving the initial PG bandgap value, and a flat midgap level (the O$_2$ energy level) appears nearby the Fermi level. These small changes in the bandgap configuration for the non-defective PG/O$_2$-H and PG/O$_2$-V systems, and their related low values of absorption energy, denote that this kind of PG lattice is not a useful nanostructure when it comes to gas sensing applications. This trend was also reported for the graphene case \cite{yan_JAP,sun_nanoscale}. The four right-most panels (PG@A/O$_2$-H, PG@A/O$_2$-V, PG@B/O$_2$-H, PG@B/O$_2$-V) show that electronic band structure of defective PG lattices upon O$_2$ adsorption are substantially impacted. As discussed above, these cases have presented higher adsorption energies. Their bandgap configurations show the following common trend: upon adsorption, the bandgap value is slightly decreased and some flat midgap levels are formed. The PG@B/O$_2$ band structures are insensitive to the O$_2$ orientation regarding the PG plane whereas the PG@A/O$_2$ ones show changes in their configuration (i.e., PG@A lattice present a degree of selectivity).        

\begin{figure*}[pos=ht]
\centering
\includegraphics[width=0.95\linewidth]{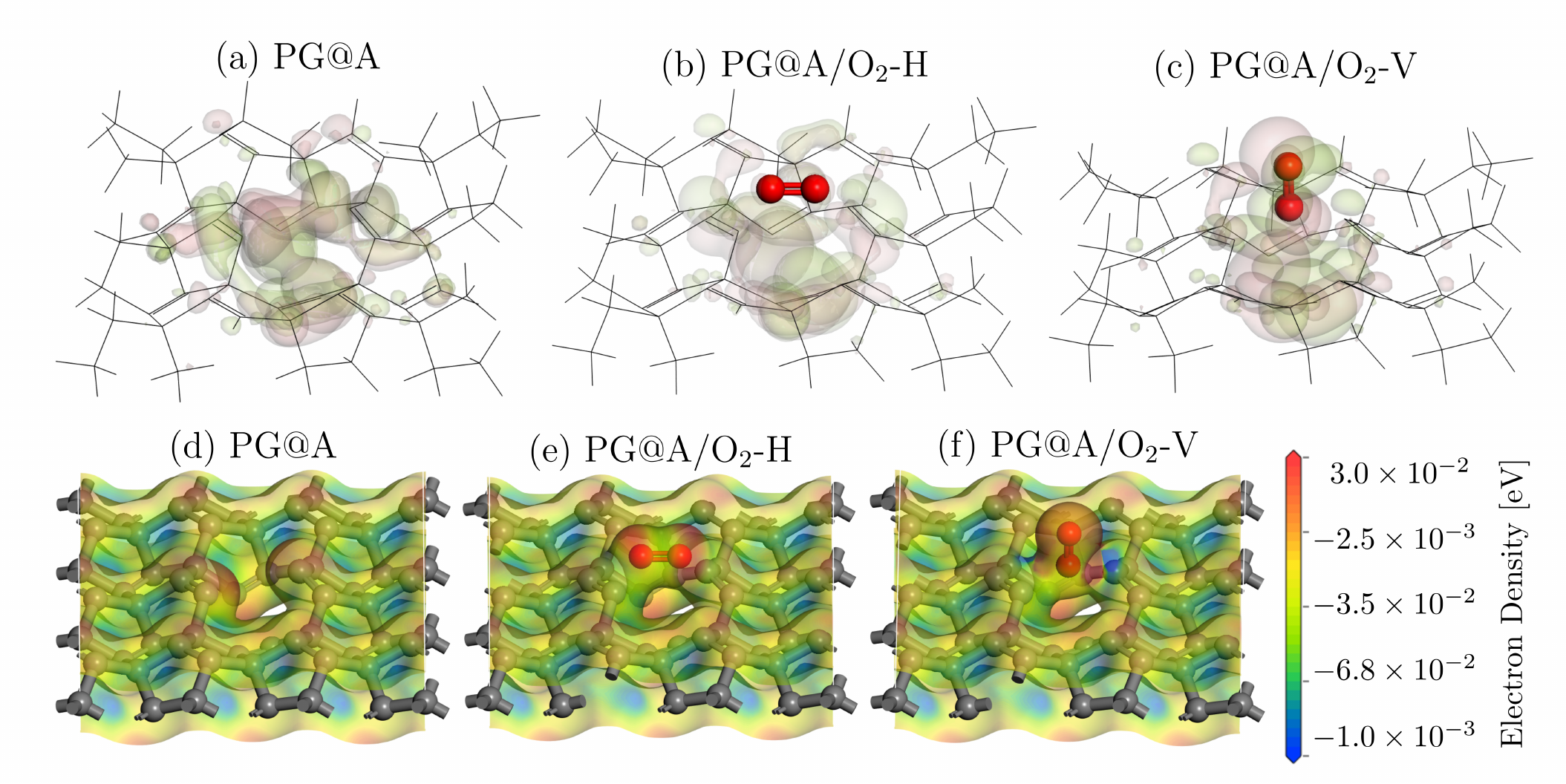}
\caption{(top panels )Electrostatic potential over the structure and (bottom panels) the Highest Occupied Molecular Orbital (HOMO) and Lowest Unoccupied Molecular Orbital (LUMO) configurations for the case of higher adsorption energy (PG@A/O$_2$). In the color scheme for the top panels, the HOMO and LUMO orbitals are represented by the green and pink spots, respectively.}
\label{fig:orbitals}
\end{figure*}

Finally, we analyze the electrostatic potential over the structure as well as the Highest Occupied Molecular Orbital (HOMO) and Lowest Unoccupied Molecular Orbital (LUMO) configurations only for the of better adsorption performance (PG@A/O$_2$). The top and bottom panels of Figure \ref{fig:orbitals} illustrate, respectively, the HOMO and LUMO orbitals and the electronic potential. The PG@A/O$_2$-H shows the better reactivity among all the systems, once the O$_2$ molecule has a considerable degree of electronegativity while the PG@A lattice presents an excess of non-bonding electrons in the vicinity of the vacancy ($sp^3$-like single-atom vacancy). This feature is illustrated by the electrostatic potential results, as shown in the bottom panels of Figure \ref{fig:orbitals}. The HOMO (green spots) and LUMO (pink spots) changed their localization upon the O$_2$ adsorption, as can be seen in the top panels of Figure \ref{fig:orbitals}. Notably, in the PG@A/O$_2$-H, part of the O$_2$ HOMO/LUMO orbital is delocalized on the PG surface. In this delocalization pattern, overlapping between the O$_2$ and PG@A orbitals takes place, which considerably enhances the interaction between. This orbitals overlapping is the main responsible in promoting the higher adsorption energy values and, consequently, the higher recovering time and changes in the electronic band structure discussed above (Figures \ref{fig:recover} and \ref{fig:bandgap}, respectively). It is worthwhile to stress that in the PG@B cases ($sp^2$-like single-atom vacancy) the bond reconstructions (see Figure \ref{fig:structures}) substantially diminishes the number of non-bonding electrons in the vicinity of the vacancy. Due to this reason, the PG@B cases presented lower adsorption energies when contrasted to the PG@A ones.      

\section{Conclusions}
\label{sec4}

In summary, we carried out DFT calculations to investigate the O$_2$ adsorption mechanism on the surface of PG lattices endowed of single-atom vacancies. Our results have revealed that the relationship between the adsorption energy and the distance between O$_2$ and PG yields typical potential energy curves. The PG/$O_2$-H, PG/O$_2$-V, PG@B/O$_2$-H, and PG@B/O$_2$-V showed a physisorption mechanism. On the other hand, the PG@A/O$_2$-H e PG@A/O$_2$-V cases present higher reactivity. The adsorption energy for the PG@A/O$_2$-H case is, at least, twice higher than all the other cases. Particularly, the adsorption process in PG@A/O$_2$-H system stands for a chemisorption mechanism. 

The recovering time values for PG@A/O$_2$-H and PG@A/O$_2$-V adsorption energy curves varied from 0.5 ps up to 2.5 ps. For the lowest absorption energy (PG@A/O$_2$-H case) $\tau=2.31$ ps whereas for the PG@A/O$_2$-V one, the highest recovering time is about 1.5 ps. These transient times were large enough to realize the interaction between O$_2$ and PG@A and they have allowed changes in the electronic properties of the PG@A lattice. The electronic band structure of defective PG lattices upon O$_2$ adsorption is substantially impacted. Their bandgap configurations show the following common trend: upon adsorption, the bandgap value is slightly decreased and some flat midgap levels are formed. The PG@B/O$_2$ band structures are insensitive to the O$_2$ orientation regarding the PG plane whereas the PG@A/O$_2$ ones show changes in their configuration, by presenting a degree of selectivity.    

In the PG@A/O$_2$-H, part of the O$_2$ HOMO/LUMO orbital is delocalized on the PG surface and overlapping between the O$_2$ and PG@A orbitals is observed, which considerably enhances the interaction between. This overlapping trend is the main responsible in promoting the higher adsorption energy values and, consequently, the higher recovering time and changes in the electronic band structure for the PG@A/O$_2$-H case.

\section*{Acknowledgements}

The authors gratefully acknowledge the financial support from Brazilian Research Councils CNPq, CAPES, and FAPDF and CENAPAD-SP for providing the computational facilities. L.A.R.J. gratefully acknowledges, respectively, the financial support from FAP-DF  and CNPq grants 00193.0000248/2019-32 and 302236/2018-0. 

\printcredits
\bibliographystyle{unsrt}
\bibliography{cas-refs}

\end{document}